\pdfoutput=1
\documentclass[12pt]{article}
\usepackage{graphicx}

\newcommand\pubnumber{SNSN-XXX-YY}
\newcommand\pubdate{\today}

\newcommand{\Tchem}{\ensuremath{T_{chem}}}
\newcommand{\DeltaptN}{\ensuremath{\Delta[P_{T},N]}}
\newcommand{\SigmaptN}{\ensuremath{\Sigma[P_{T},N]}}
\newcommand{\Phipt}{\ensuremath{\Phi_{p_{T}}}}
\newcommand*{\pT}{\ensuremath{p_{\mathrm{T}}}}
\newcommand*{\mub}{\ensuremath{\mu_{B}}}
\newcommand{\agev}{\mbox{$A$~GeV}}               

\newcommand{\meanpt}{\mbox{$\langle p_{T} \rangle$}}
\newcommand*{\roots}{\ensuremath{\sqrt{s_{_{NN}}}}}

\newcommand{\PhiAB}{\ensuremath{\Phi_{AB}}}
\newcommand{\Phipipp}{\ensuremath{\Phi_{\pi(p+\bar{p})}}}
\newcommand{\PhipiKK}{\ensuremath{\Phi_{\pi(K^{+}+K^{-})}}}
\newcommand{\PhippKK}{\ensuremath{\Phi_{(p+\bar{p})K}}}

\newcommand{\PhipK}{\ensuremath{\Phi_{pK^{+}}}}

\def\napoli{Institute of Physics \\
Jan Kochanowski University, Swietokrzyska 15, 25-406 Kielce, POLAND}
\def\support{\footnote{Work supported by the Polish National Science Centre under contracts 
          on the basis of decisions no. DEC-2011/03/B/ST2/02617, DEC-2012/04/M/ST2/00816}}

\def\Title#1{\begin{center} {\Large #1 } \end{center}}
\def\Author#1{\begin{center}{ \sc #1} \end{center}}
\def\Address#1{\begin{center}{ \it #1} \end{center}}

\newcommand\pubblock{\rightline{\begin{tabular}{l} \pubnumber\\
         \pubdate  \end{tabular}}}
\newenvironment{Abstract}{\begin{quotation}  }{\end{quotation}}
\newenvironment{Presented}{\begin{quotation} \begin{center} 
             PRESENTED AT\end{center}\bigskip 
      \begin{center}\begin{large}}{\end{large}\end{center} \end{quotation}}

\begin{document}
\begin{titlepage}
\pubblock

\vfill
\Title{Recent results from the search for the critical point of strongly interacting matter at the CERN SPS}
\vfill
\Author{ Grzegorz Stefanek\support \\for the NA49 and NA61/SHINE Collaborations}
\Address{\napoli}
\vfill
\begin{Abstract}
Recent searches at the CERN SPS for evidence of the critical point of strongly interacting matter are discussed.
Experimental results on theoretically expected signatures, such as event-to-event fluctuations of the particle multiplicity and the average transverse momentum 
as well as intermittency in particle production are presented.
\end{Abstract}
\vfill
\begin{Presented}
XXXIV Physics in Collision Symposium \\
Bloomington, Indiana,  September 16--20, 2014
\end{Presented}
\vfill
\end{titlepage}
\def\thefootnote{\fnsymbol{footnote}}
\setcounter{footnote}{0}

\section{Introduction}

The exploration of the phase diagram of strongly interacting matter, particularly the search for a phase transition from hadronic to partonic degrees of freedom and possibly a critical endpoint, is one of the most challenging tasks in present heavy ion physics.

\begin{figure}[h!]
\vspace{-0.4cm}
\begin{center}
\includegraphics[width=5.0cm,trim=0.0cm 1.8cm 10.8cm 1.4cm,clip=true]{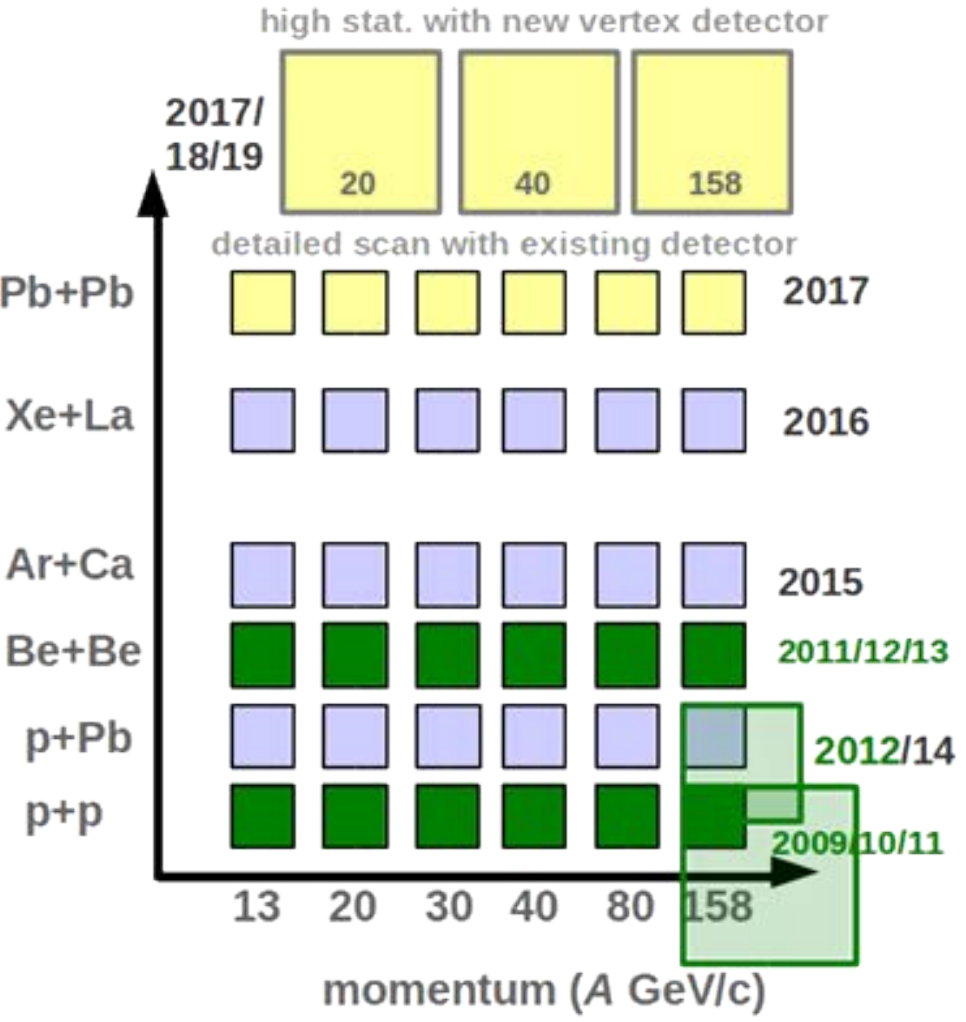}
\includegraphics[width=7.8cm,trim=0.0cm 0.0cm 0.0cm 0.0cm,clip=true]{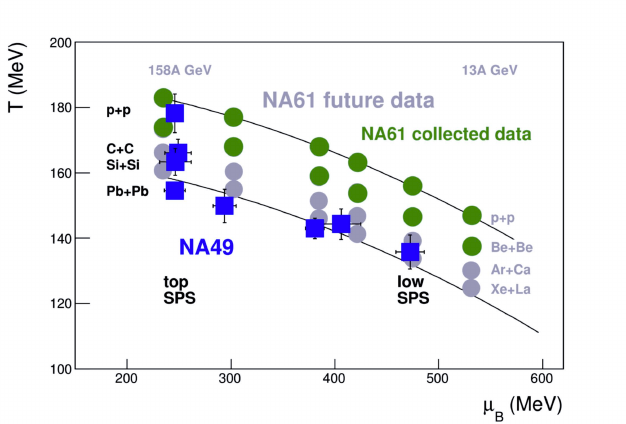}
\end{center}
\vspace{-0.6cm}
\caption{Left: Data sets collected (green) and planned to be recorded by NA61/SHINE within (light blue) and beyond (yellow) the approved ion program. Right: The planned scan of the phase diagram by varying collision energy (\mub) and size of colliding nuclei ($T$). Chemical freeze-out points are taken from~\cite{Becattini:2005xt} and parametrizations therein.}
\label{fig:programme}
\vspace{-0.4cm}
\end{figure}

	NA49 data on inclusive hadron production indicate that the onset of deconfinement in central Pb+Pb collisions is located at 30\agev~beam energy(\roots=7.7~GeV). It is mainly based on the observation of norrow structures in the energy dependence of hadron production in central Pb+Pb collisions which are not observed in elementary interactions~\cite{Alt:2007aa,Gazdzicki:2004ef}. 
The NA61/SHINE experiment~\cite{NA61:webpage} continues the ion program of NA49 with the main aim of searching for the critical point and studying in detail the onset of deconfinement by performing a two dimensional scan of the phase diagram in $T$ and \mub . This is achieved by varying collision energy (13$A$-158\agev) and size of the colliding systems. The data sets recorded by both experiments and those planned to be recorded by NA61/SHINE for the scan of the phase diagram are presented in Fig.~\ref{fig:programme}. Chemical freeze-out points in  Fig.~\ref{fig:programme}(right) are taken from~\cite{Becattini:2005xt}. The presence of the predicted critical point is expected to lead to an increase of event-by-event fluctuations of many observables~\cite{Stephanov:1999zu,Berdnikov:1999ph} provided that the freeze-out of the measured hadrons occurs close to its location in the phase diagram and the evolution of the final hadron phase does not erase the fluctuation signals. The NA61/SHINE experiment looks for a maximum of fluctuations as experimental signature for the critical point.

\section{Event-by-event fluctuations}
\label{Sect:EbE fluctuations}

\begin{figure}[h!]
\vspace{-0.2cm}
\begin{center}
\includegraphics[width=7.4cm,trim=0.0cm 0.0cm 0.0cm 1.02cm,clip=true]{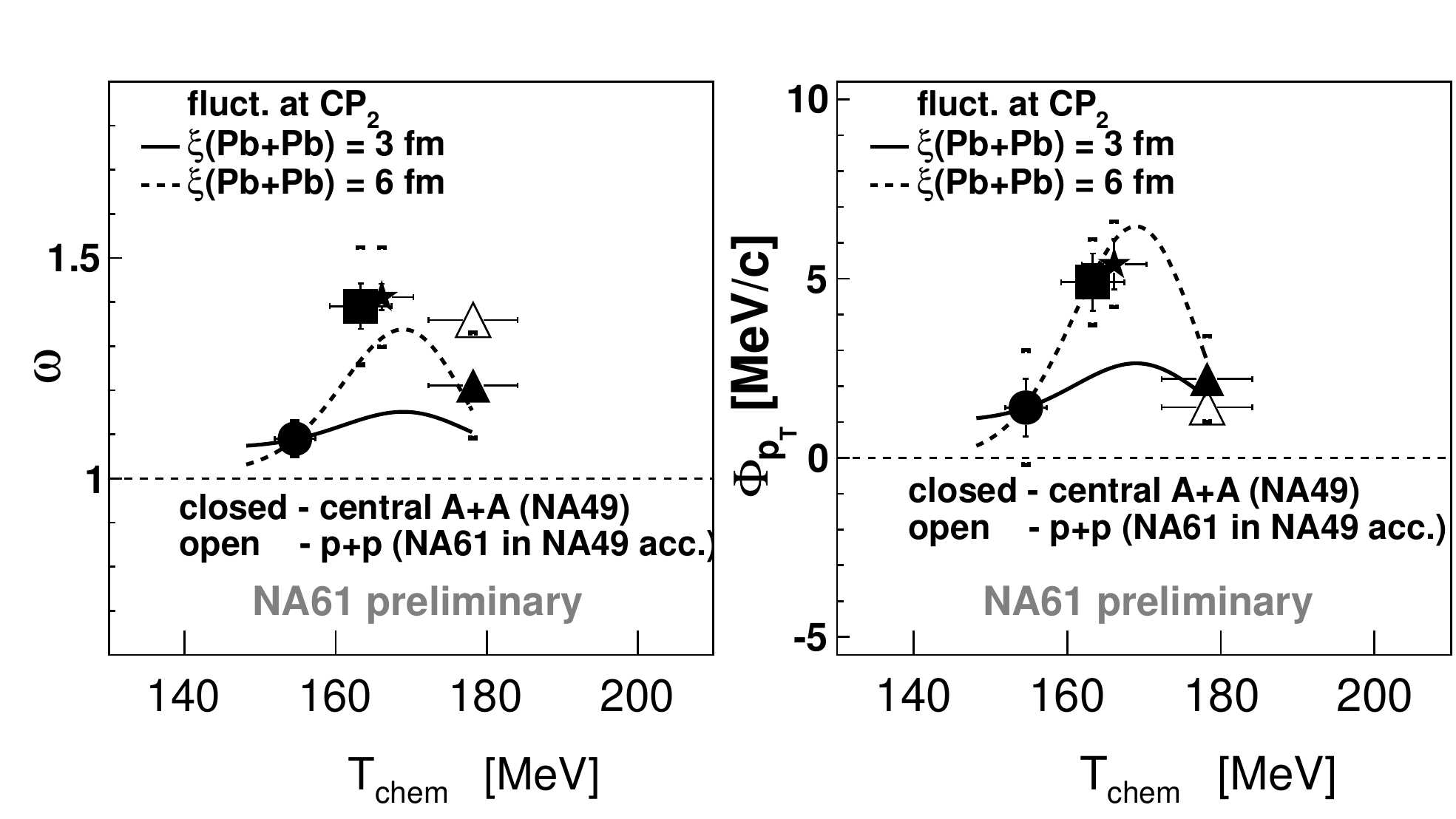}
\includegraphics[width=7.4cm,trim=0.0cm 0.0cm 0.0cm 1.02cm,clip=true]{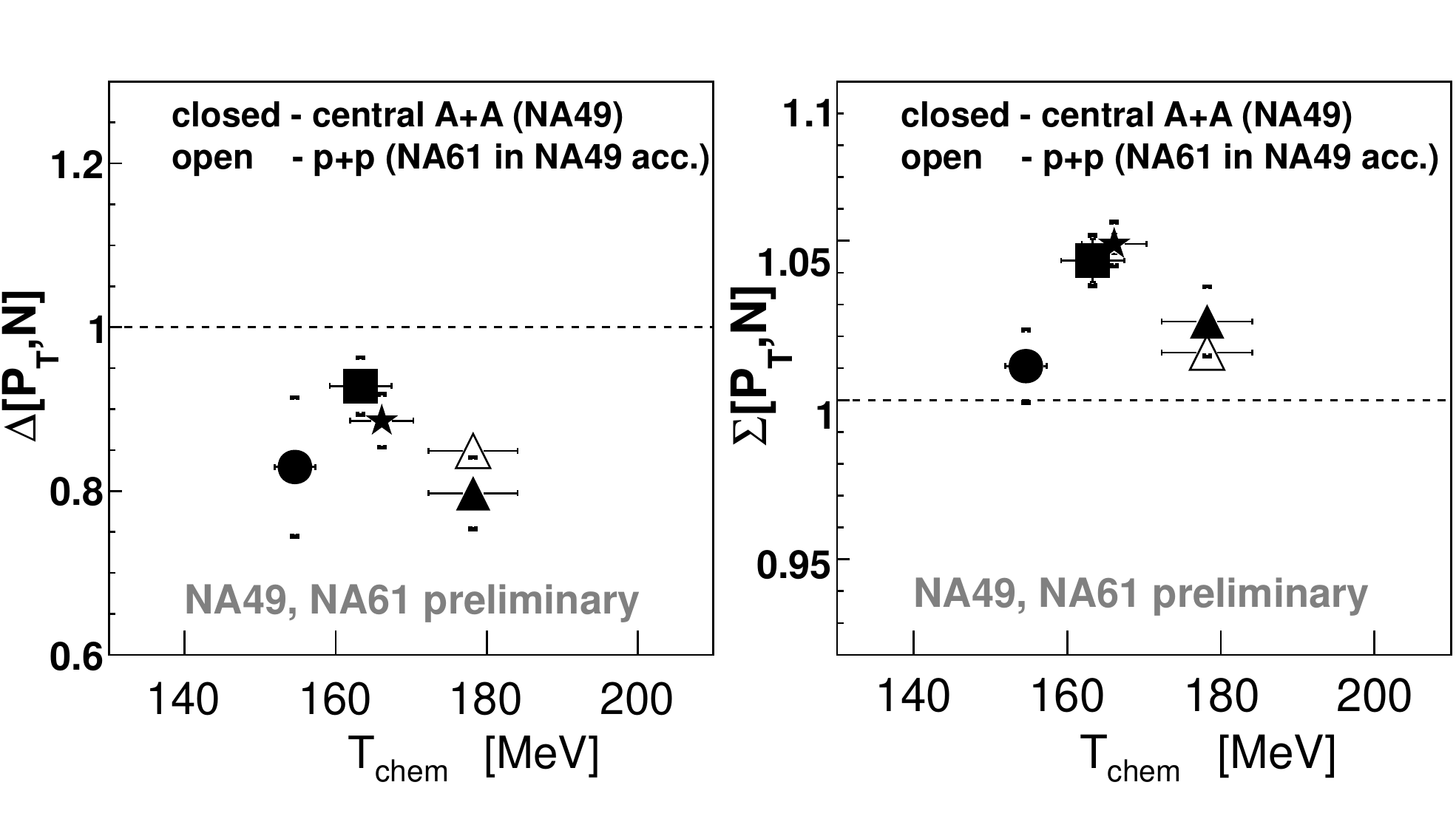}
\vspace{-0.4cm}
\caption{The dependence of  the scaled variance of multiplicity fluctuations $\omega$ and the \Phipt, \DeltaptN, \SigmaptN~measures on \Tchem~(system size) for charged hadrons in central AA collisions and pp interactions. Curves indicate the estimated effects of the critical point ($T$=178MeV, \mub =250MeV) for two values of the correlation length $\xi$.}
\label{fig:EbE_vs_temp}
\end{center}
\vspace{-0.6cm}
\end{figure}

    In the NA49 and NA61/SHINE experiments event-by-event fluctuations of the particle multiplicity are measured by the scaled variance $\omega$ which is an intensive measure (independent of interaction volume) and defined as
\begin{equation}
\omega = \omega \left[N\right] = \frac{\langle N^{2} \rangle - \langle N \rangle ^{2}}{\langle N \rangle}
\label{Eq:omega}
\end{equation}
The transverse momentum fluctuations are measured by the $\Phi$ quantity proposed in~\cite{Gazdzicki:1992ri} which was also succesfully employed by the NA49 experiment to study charge and azimuthal angle fluctuations~\cite{Alt:2004ir,Cetner:2010vz}. Recently new strongly intensive fluctuation measures (independent of volume and volume fluctuations) $\Delta$ and $\Sigma$ were proposed in~\cite{Gazdzicki:2013ana}.  Whereas \SigmaptN~is closely related to \Phipt~the quantity \DeltaptN~is sensitive to fluctuations of \meanpt~and $N$ in another combination and thus these two measures can be sensitive to different physics effects. 
Fig.~\ref{fig:EbE_vs_temp} shows the dependence of multiplicity fluctuations quantified by the scaled variance $\omega$ and fluctuations of the average transverse momentum measured by \Phipt~as well as \DeltaptN~ and \SigmaptN~as a function of chemical freeze-out temperature \Tchem~(obtained from statistical model fits ~\cite{Becattini:2005xt}) in p+p interactions and central C+C, Si+Si and Pb+Pb collisions. All quantities show a possible maximum of fluctuations at \Tchem=160-170 MeV reached in collisions of medium-size nuclei. This agrees with expactations for the critical point represented by curves for $\omega$ and \Phipt~\cite{Grebieszkow:2009jr}. 

\begin{figure}[h!]
\vspace{-0.18cm}
\begin{center}
\includegraphics[width=7.4cm,trim=0.2cm 0.0cm 0.0cm 1.02cm,clip=true]{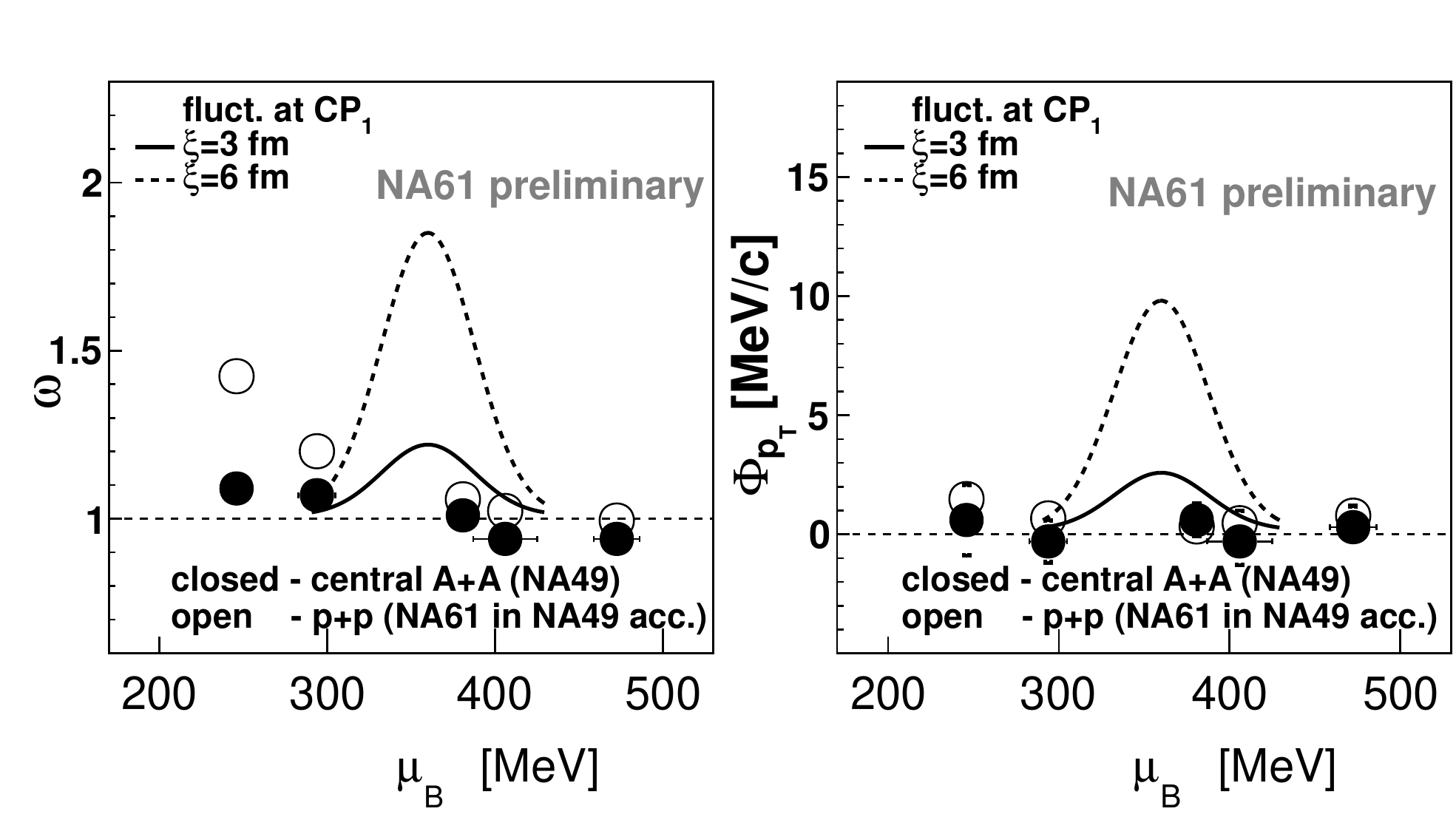}
\includegraphics[width=7.4cm,trim=0.2cm 0.0cm 0.0cm 1.02cm,clip=true]{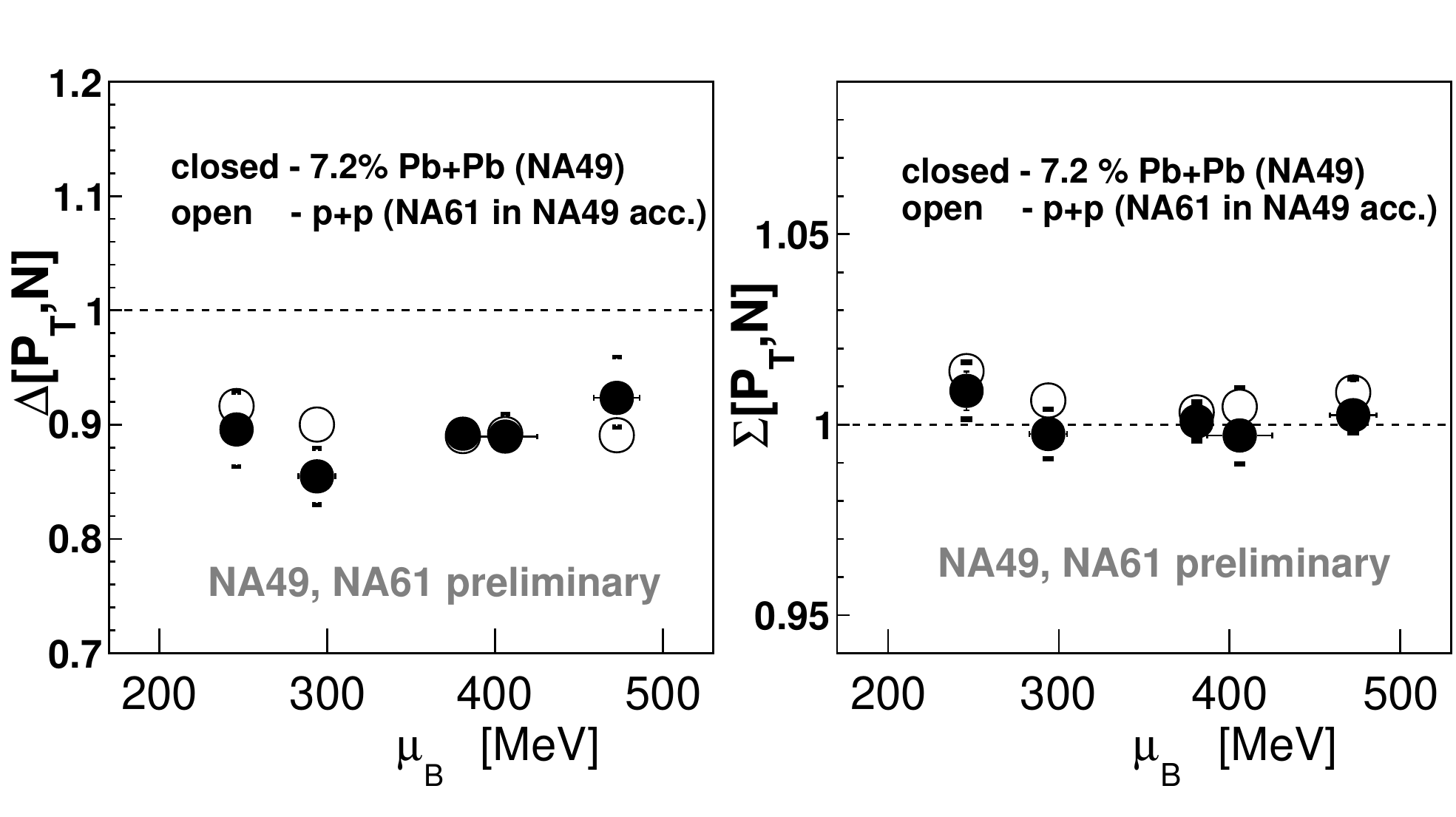}
\vspace{-0.4cm}
\caption{The dependence of the scaled variance of multiplicity fluctuations $\omega$ and the \Phipt, \DeltaptN, \SigmaptN~measures on \mub  (collision energy) for charged hadrons in central PbPb collisions and pp interactions. Curves indicate the estimated effects of the critical point ($T$=147MeV, \mub =360MeV) for two values of the correlation length $\xi$.}
\label{fig:EbE_vs_miub}
\end{center}
\vspace{-0.8cm}
\end{figure}

The dependence of all fluctuation measures on \mub (obtained from statistical model fits [4]) in p+p interactions (NA61) and central Pb+Pb collisions (NA49) is shown in Fig.~\ref{fig:EbE_vs_miub}. Here the data don't indicate a maximum as might be expected for the critical point (curves for $\omega$ and \Phipt~\cite{Grebieszkow:2009jr}).

A new identification procedure called identity method~\cite{Gazdzicki:2011xz,Gorenstein:2011hr} allowed to measure the energy dependence of fluctuations of identified proton, kaon and pion multiplicities in p+p and Pb+Pb collisions (plots not shown because of space limitations). As in the case of charged particle multiplicites no indication of the critical point was found.   

    Chemical (particle type) fluctuations for two different particle types $A$ and $B$ were analysed using the \PhiAB~measure (see ~\cite[Eq.6,7]{Gorenstein:2011vq}).

\begin{figure}[h!]
\vspace{-0.2cm}
\begin{center}
\includegraphics[width=5.0cm,trim=0.44cm 0.3cm 2.3cm 0.5cm,clip=true]{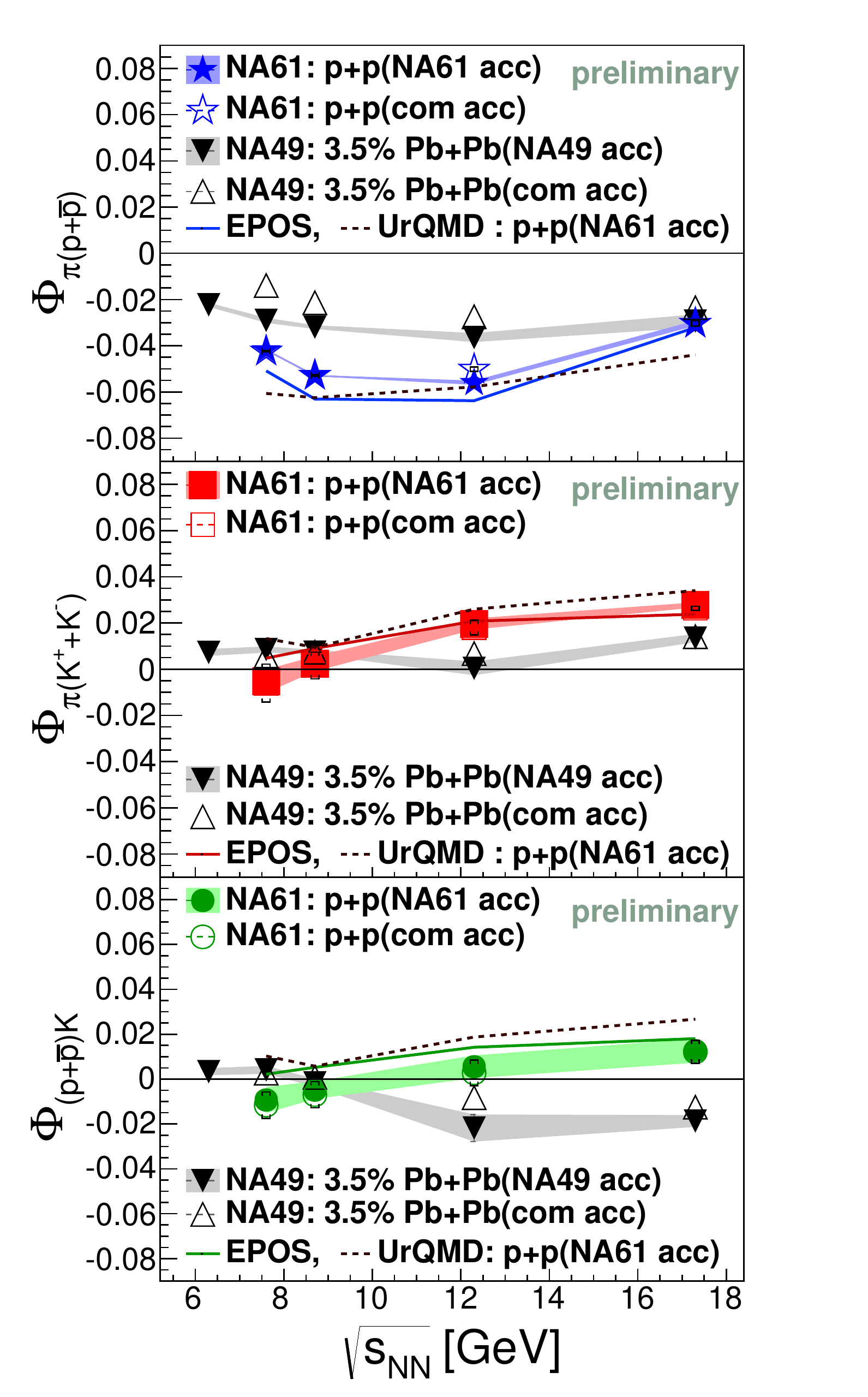}
\includegraphics[width=5.0cm,trim=0.45cm 0.3cm 2.3cm 0.5cm,clip=true]{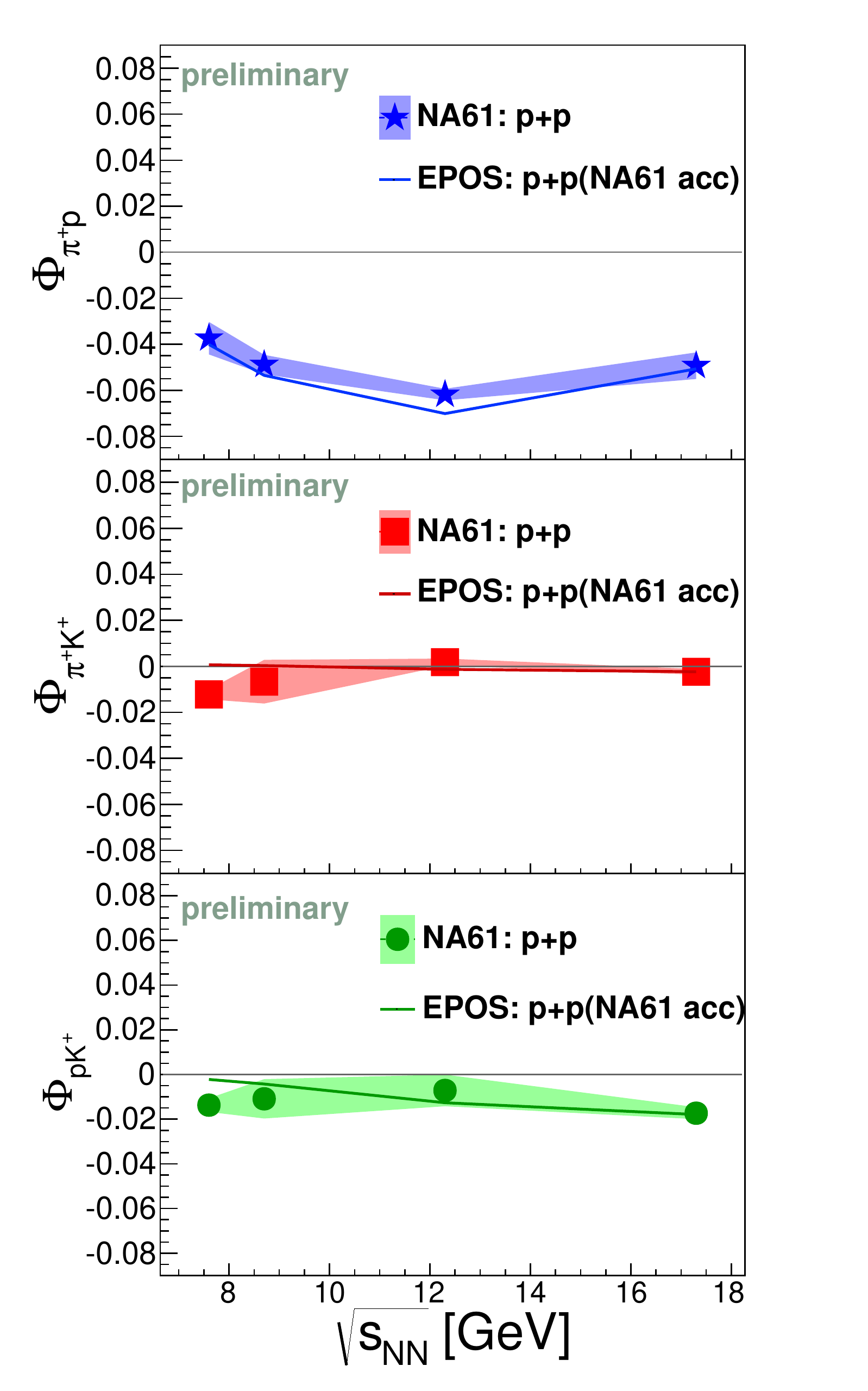}
\vspace{-0.2cm}
\caption{Chemical (particle type) fluctuations in inelastic pp interactions at 30, 40, 80, and 158GeV/c. NA61 pp data are compared to those obtained by NA49 in PbPb collisions~\cite{Rustamov:2013oza}. See~\cite{acceptance:edms} for NA49, NA61 acceptance and common phase space region.}
\label{fig:chemical_fluct}
\end{center}
\vspace{-0.4cm}
\end{figure} 

      Fig.~\ref{fig:chemical_fluct} shows two particle chemical fluctuations as a function of interaction energy. The fluctuations cannot be corrected for the limited acceptance thus the results are presented both for the NA49, NA61 acceptance and a common NA49/NA61 phase space region. The left panel of Fig.~\ref{fig:chemical_fluct} shows combinations of both charges while the right one shows only positively charged particles. The values of \Phipipp~are negative for all studied energies. This is most probably due to charge conservation and resonance decays. In p+p collisions \PhipiKK~is higher than zero probably due to strangeness conservation. The slight increase with energy for p+p interactions is not observed for Pb+Pb collisions. For both systems \PhippKK~crosses zero at medium SPS energies. Finally, we observe no significant energy dependence of \PhipK in the NA61 data. The EPOS and UrQMD model predictions reproduce p+p data resonably well. 

\section{Density fluctuations of protons and low-mass $\pi^{+}\pi^{-}$ pairs}
\label{Sect:Intermittency}

\begin{figure}[h!]
\vspace{-0.4cm}
\begin{center}
\includegraphics[width=6.4cm,trim=0.5cm 0.3cm 0.0cm 0.7cm,clip=true]{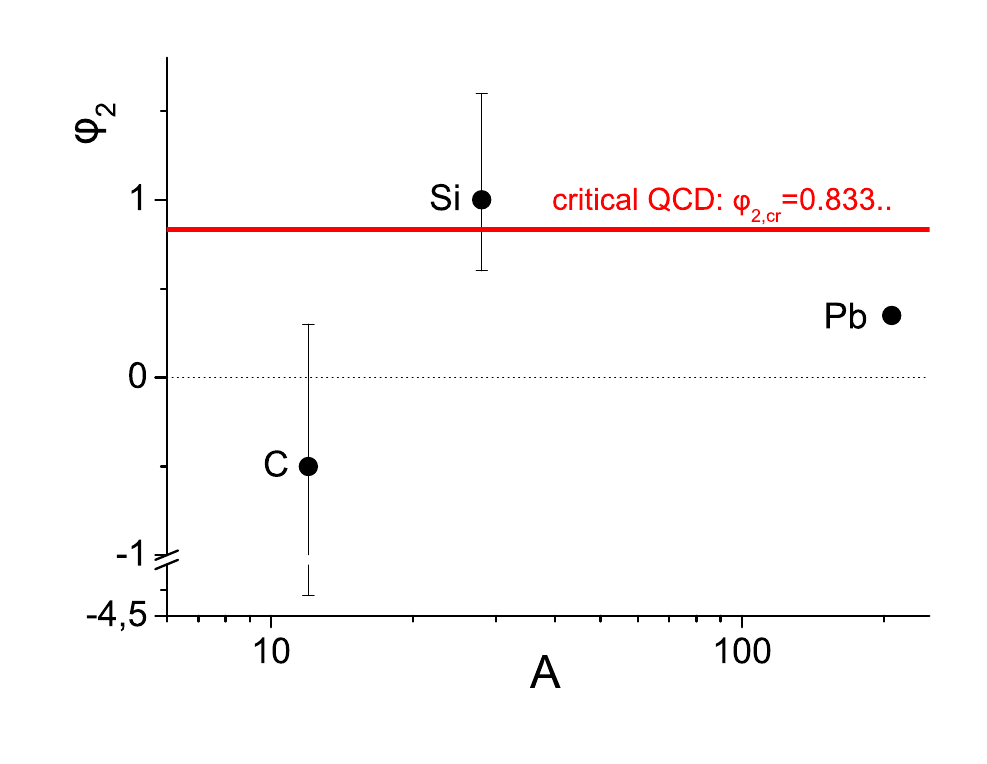}
\includegraphics[width=6.0cm,trim=0.5cm 0.5cm 1.5cm 0.7cm,clip=true]{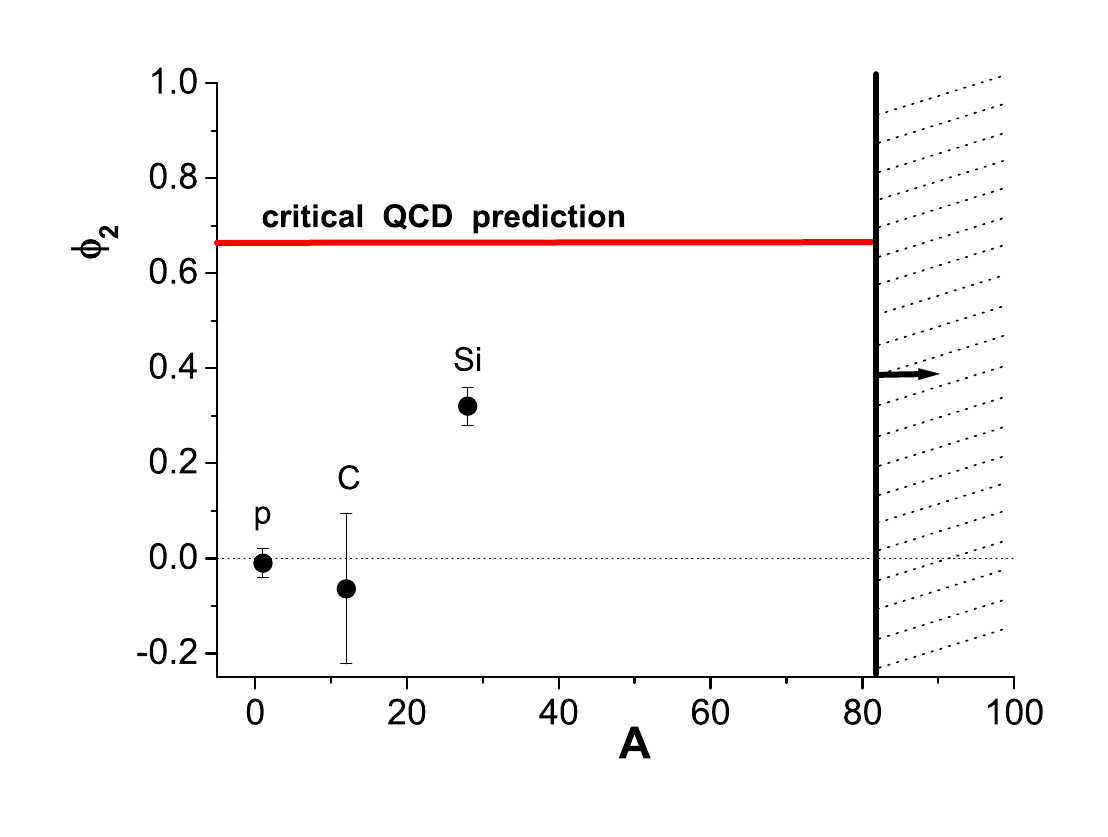}
\vspace{-0.5cm}
\caption{Exponent $\phi_{2}$ obtained from power-law fits to corrected second scaled factorial moments of protons (left) and low-mass $\pi^{+}\pi^{-}$ pairs (right) for several collision systems at 158AGeV. }
\label{fig:critical_fluct}
\end{center}
\vspace{-0.8cm}
\end{figure}

Theoretical investigation~\cite{Antoniou:2000ms2005am} predicts near the critical point the appearance of local density fluctuations  for  protons~\cite{Hatta:2003wn} and low-mass $\pi^{+}\pi^{-}$ pairs of power-law nature with known critical exponents~\cite{Antoniou:2006zb}. This can be studied by the intermittency analysis method in transverse momentum space using second factorial moments $F_{2}(M)$, where M is the number of subdivisions in each $\pT$ direction. After combinatorial background subtraction the exponents $\phi_{2}$ are obtained from a power-law fit to the corrected moments $\Delta F_{2}(M) \approx M^{2\phi_{2}}$. The resulting values of $\phi_{2}$ obtained for central C+C, Si+Si and Pb+Pb collisions at 158\agev~are plotted in Fig.~\ref{fig:critical_fluct}. Remarkably, $\phi_{2}$  seems to reach a maximum for Si+ Si collisions  which is consistent with the theoretical expectations for the critical point.

\end{document}